\documentclass[aps,prl,twocolumn,showpacs,floatfix]{revtex4-1}
\usepackage{graphics}
\usepackage{graphicx}
\usepackage{amssymb,amsmath}

\usepackage{siunitx}
\usepackage{lineno}

\usepackage{xcolor}
\usepackage{pgf, tikz}
\usepackage{pgfplots}
\pgfplotsset{compat=1.3}

\definecolor{bmblue}{rgb}{0.8,0.8,1}
\definecolor{jhred}{rgb}{1, 0.8, 0.8}
\definecolor{darkgreen}{rgb}{0, 0.5, 0}

\usepackage{soul}
\soulfont{\cite}{1}
\setstcolor{red}

\usepackage{booktabs}

\begin{document}

\title{Eigenmodes in the long-time behavior of a coupled spin system measured with nuclear magnetic resonance}



\author{Benno Meier}
\affiliation{University of Leipzig, Faculty of Physics and Earth Science, Linn\'{e}strasse 5, 04103 Leipzig, Germany}

\author{Jonas Kohlrautz}
\affiliation{University of Leipzig, Faculty of Physics and Earth Science, Linn\'{e}strasse 5, 04103 Leipzig, Germany}

\author{J\"{u}rgen Haase}
\affiliation{University of Leipzig, Faculty of Physics and Earth Science, Linn\'{e}strasse 5, 04103 Leipzig, Germany}


\date{\today}
\begin{abstract}
  The many body quantum dynamics of dipolar coupled nuclear spins $I = 1/2$ on an otherwise isolated cubic lattice are studied with nuclear magnetic resonance (NMR). By increasing the signal-to-noise ratio by two orders of magnitude compared with previous reports for the free induction decay (FID) of $^{19}$F in CaF$_2$ we obtain new insight into its long-time behavior. We confirm that the tail of the FID is an exponentially decaying cosine, but our measurements reveal a second decay mode with comparable frequency but twice the decay constant. This result is in agreement with a recent theoretical prediction for the FID in terms of eigenvalues for the time evolution of chaotic many-body quantum systems.
\end{abstract}




\maketitle

One of the simplest NMR experiments concerns the flipping and recording of the nuclear magnetization, i.e., the measurement of the free induction decay (FID). Quite to the contrary, rigorous theory to calculate this decay is often lacking since large numbers of nuclear spins are interacting, similar to the situation in electronic magnetism with exchange coupled electronic spins. Often, one can only estimate the short-time behavior of the FID, but is unable to find an analytical expression for the entire decay. If the FID is limited by the life-time of the nuclear levels due to a coupling to a thermal bath, the "lattice", the decay is simply exponential and given by the spin-lattice relaxation time $T_1$. This can be the case in liquids where rapid motion averages, e.g., the inter-nuclear magnetic dipole-dipole interaction, but, at the same time, couples the nuclear spins to the large thermal bath of motional degrees of freedom. In solids, where the spin-lattice relaxation time is often rather long, the magnetic dipole interaction leads to a rapid decay of the FID leaving the spin system far away from thermal equilibrium for times of the order of $T_1$. In this case, where the FID or other spin coherences decay according to the time-evolution of the acting Hamiltonian, one expects on general grounds the decay not to be exponential. However, that is what is often observed with experiments \cite{lowe_free-induction_1957,engelsberg_free-inducation-decay_1974,sorte_long-time_2011,morgan_universal_2008}.

An ideal system for the investigation of dipole coupled nuclear spins is CaF$_2$ where spin 1/2 fluorine nuclei are located on a simple cubic lattice (the low abundance, small moment Ca spins can be neglected).
Clean crystals are easily available with a nuclear spin-lattice relaxation time $T_1$ of minutes, which practically leaves only one relevant time scale set by the dipolar coupling ($\sim \SI{20}{\micro\second}$).  Given the simple nature of the material and the challenging physics, there has been persistent interest in CaF$_2$ since the early days of magnetic resonance \cite{van_vleck_dipolar_1948,lowe_free-induction_1957,engelsberg_free-inducation-decay_1974,slichter_principles_2010,cowan_nuclear_2005,abragam_principles_1983,cho_multispin_2005,sorte_long-time_2011}.

While most experimental and theoretical work focussed on the short-time behavior of the dynamics \cite{van_vleck_dipolar_1948,lowe_free-induction_1957,betsuyaku_lowe-norberg_1970,canters_numerical_1972,jensen_sixth_1973}, more recent research concerns the \textit{long-time} behavior of NMR signals \cite{borckmans_long-time_1968,pastawski_nuclear_2000,fine_long-time_2004,fine_long-time_2005}. In particular in recent work of Fine \cite{fine_long-time_2004,fine_long-time_2005}, the long-time behavior of the system is addressed based on the notion of microscopic quantum chaos.

The dynamics of a lattice of \textit{classical} spins can be described by a set of angles $\{\phi_i, \theta_i \}$ where the $i$-th pair of coordinates describes the orientation of the $i$-th spin with its tip on the surface of a sphere. In case of a dipole coupled system the equations of motion controlling the time evolution of the system are nonlinear. For a large number of spins this eventually leads to a phase space that is dominated by chaotic regions. While a Markovian description of such a system is usually applicable only for times much larger than the mean free time, this is not true for ensemble averaged quantities \cite{krylov_works_1979}. An ensemble of spins can therefore be described in terms of Brownian motion as soon as the spin system has lost memory of its initial configuration. Thus, the long-time behavior of the ensemble averaged quantities can be obtained by solving a correlated diffusion equation on a spherical surface \cite{fine_long-time_2004}.

This approach is to be contrasted with chaotic dynamics of the magnetization in liquids. Here, the local field generated by the distant dipole field as well as the radiation damping field may cause a \textit{turbulent} behavior of the macroscopic magnetization \cite{lin_resurrection_2000,jeener_dynamical_1999}. While chaotic dynamics in liquids are discussed in an entirely classical picture, the microscopic diffusion concept for solids has recently been generalized to quantum spins by Fine \cite{fine_long-time_2004,fine_long-time_2005}. He predicted a \textit{universal} long-time behavior of the FID and the decay after a spin echo.

For a system of spins $1/2$ on any Bravais lattice with the dipolar Hamiltonian
\begin{equation}
  \label{eq:dipolarHamiltonian}
  \mathcal{H}_{\mbox{\tiny dip}} = \frac{\gamma_n^2 \hbar^2}{2} \sum_{j,k}^{N} \left [ \frac{\boldsymbol{I}_j \boldsymbol{I}_k}{r_{jk}^3}  - \frac{3 (\boldsymbol{I}_j \cdot \boldsymbol r_{jk} ) (\boldsymbol{I}_k \cdot \boldsymbol{r}_{jk})}{r_{jk}^5} \right ],
\end{equation}
in a strong magnetic field (so that only the secular part has to be retained) the long-time behavior is given by
$G\left( t \right)~\simeq~\sum\nolimits_\sigma  {{e^{ - {\gamma _\sigma }t}}\cos \left( {\omega_\sigma t + \phi_\sigma } \right)}$.
This reduces to a single mode after sufficiently long times, i.e.,
\begin{equation}
  \label{eq:longTimeTwo}
  G(t) \simeq e^{ - {\gamma _1 }t}\cos \left( {\omega_1 t + \phi} \right).
\end{equation}
In related work, the long-time behavior is referred to as that part of the FID that is determined by the slowest mode only. In the experiment, this description holds after the second slowest mode has disappeared in the noise. Thus the beginning of the long-time behavior would be set by the noise level. In this work we therefore consider the long-time-behavior to correspond to that part of the FID where chaos has developed, i.e., the part that is given as the sum of all modes. The number of \textit{detectable} modes is set again by the signal-to-noise ratio (c.f. Fig.~S1 in supplement {\cite{meier_see_2012}}).

In the above conjecture, the decay modes in $G(t)$ follow from a set of solutions of a correlated surface diffusion equation in the finite parameter space of single spin variables: $f_{\sigma}(t,x) = \mathrm{e}^{- \lambda_{\sigma} t} u_{\lambda_\sigma}(x)$, where $\lambda_\sigma = \gamma_\sigma + {\rm i} \omega_\sigma$ is an in general complex eigenvalue of the integro-differential operator associated with the modified diffusion equation and $u_\lambda(x)$ its corresponding eigenfunction. In particular, the conjecture proposes that $\lambda_\sigma$ does not depend on the initial configuration of the spin system. It was argued by Fine \cite{fine_long-time_2004} that due to the Markovian nature of ensemble averages \cite{krylov_works_1979} one expects not only the decay constant of the slowest mode ($\sigma=1$) to be of the order of the inverse decay time, i.e., $\gamma_1 \sim 1/\tau$, but also the difference to the second slowest exponent should be of the order of $1/\tau$, i.e., $\gamma_2  - \gamma_1 \sim \gamma_1$. There are no predictions for~$\omega_2$.

The early work of Engelsberg and Lowe {\cite{engelsberg_free-inducation-decay_1974}} is in reasonable agreement with this theory, but inconclusive with regard to the predictions above, which triggered new experiments with hyperpolarized polycrystalline solid xenon for better signal-to-noise ratio {\cite{morgan_universal_2008}} and CaF$_2$ {\cite{sorte_long-time_2011}}. The experiments show that the long-time behavior is indeed universal since it does not depend on various initial preparations of the spin system, in agreement with Fine's theory. However, only a single mode $\lambda_1$ in the exponential decay was found. In solid xenon, the isotropic averaging over all crystal orientations makes the direct observation of an isolated second mode unlikely {\cite{fine_asymptotic_2012}}.

\begin{figure}[t]
  \includegraphics{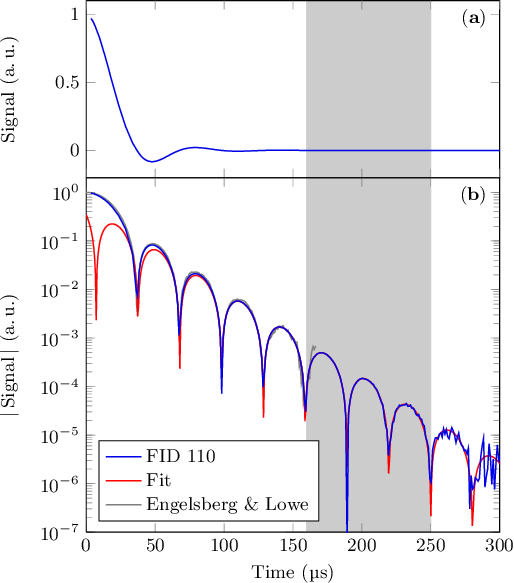}
  \caption{(\textbf{a}) Free Induction Decay of $^{19}$F in a CaF$_2$ single crystal (blue line) for $B$ parallel (110) at \SI{7.06}{T}. Note that the real part of the FID changes sign at each of the minima in the logarithmic plot (\textbf{b}). The first \SI{20}{\micro\second} of the FID where measured using a spin-locking technique, c.f. \cite{vollmers_method_1978}. The time intervals 20 to \SI{57}{\micro\second} and 58 to \SI{87}{\micro\second} where measured at \SI{20}{K} with 25 and \SI{12}{dB} attenuation in front of the preamp, respectively. The tail of the FID was measured without attenuation. The signal was corrected for the inhomogeneous decay as measured by C$_6$F$_6$ as well as for a small residual offset (\SI{700}{Hz}) that was precisely determined from the time dependence of the signal's phase. The data is in good agreement with Engelsberg and Lowe's measurements but has nearly three orders of magnitude better signal-to-noise ratio. Due to the higher signal-to-noise ratio, we were able to fit the FID to Eq.~(\ref{eq:longTimeTwo}) in the time frame 160 to \SI{250}{\micro\second} (shaded in gray).}
   \label{fig:fid}
\end{figure}

Here, we report on new experiments on CaF$_2$ that, due to the increase in signal-to-noise ratio by two orders of magnitude over previously reported experiments, reveal a second decay mode in agreement with Fine's predictions \cite{fine_long-time_2004,fine_long-time_2005}. Therefore, our results favor his theory over other theories using a memory function approach \cite{borckmans_long-time_1968}, since these predict only a single mode.

The $5\times 5 \times 10$ \si{\milli\meter\cubed} CaF$_2$ crystal was obtained from Mateck, Germany. Impurities are stated to be below {2~ppm}. The crystal was oriented by X-ray diffraction. A home-built NMR probe was used in order to align the crystal's axes $\left ( 110 \right ) $ or $\left (100 \right ) $ parallel to the applied magnetic field (\SI{7.06}{T}). The probe was set to a resonance frequency of \SI{283.383}{MHz} and operated at \SI{20}{K}. The quality factor $Q$ of the resonance circuit was \num{240}, the $\pi/2$ pulse length was \SI{5}{\micro\second}. The $T_1$ at \SI{20}{K} was determined to be \SI{76}{\second}. 

\begin{figure*}[t]
  \includegraphics{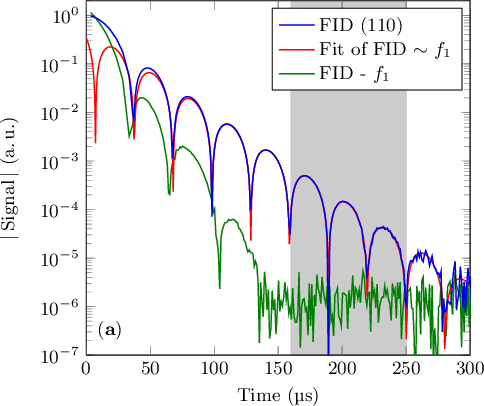}\hfill\includegraphics{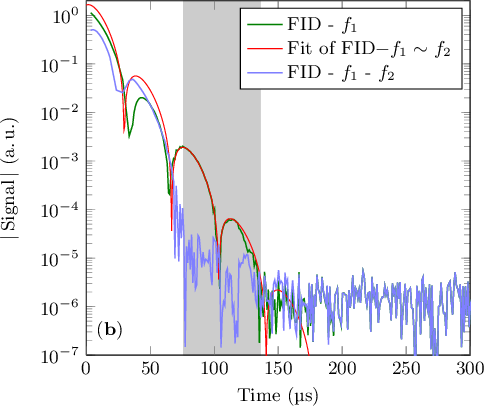}
  \caption{Extraction of two decay modes from the FID for $B$ parallel to (110). (\textbf{a}) After subtracting the fit or first Mode $f_1$ as determined by fitting the tail of the FID (c.f. Fig. \ref{fig:fid}) to Eq.~(\ref{eq:longTimeTwo}) in the time interval {\SI{160}{\micro\second}} to {\SI{250}{\micro\second}}, a second mode $f_2$ becomes apparent. This mode is replotted in (\textbf{b}) along with a fit according to Eq. (\ref{eq:longTimeTwo}) in the time interval 76 to \SI{136}{\micro\second} (shaded in gray). The fit reveals that the decay constant $\gamma_2$ of the second mode is about a factor of two larger as compared to the first mode while the obtained value for $\omega$ differs rather slightly from the first mode's one. The light blue line is left after subtracting both modes from the FID and reveals that within experimental resolution the FID is accurately described by $f_1 + f_2$ for times greater than \SI{75}{\micro\second} or for a decay over four orders of magnitude.}
  \label{fig:fidTwoModes}
\end{figure*}

With typically 80 scans our maximum signal to noise was about $10^6$. In order to record the signal, various attenuations were used to prevent the preamp from saturating and to allow for an appropriate load of the 16 bit digitizer. Additionally, the first \SI{20}{\micro\second} were measured using a spin locking sequence \cite{vollmers_method_1978}. Finally, the tail of the FID was measured with the variable attenuator set to zero but still infront of the preamp to minimize additional phase shifts. The composition of the FID is described in greater detail in the supplement material {\cite{meier_see_2012}}.

In order to determine and remove the influence of the inhomogeneous magnetic field on the decay, we measured a $^{19}$F FID on a C$_6$F$_6$ sample (intrinsic $T_2 \gg \SI{1}{ms}$) of about the same volume and geometry as the crystal. On the time-scale of the CaF$_2$ FID the $^{19}$F decay could be approximated by a Gaussian with a standard deviation of \SI{632(3)}{\micro\second}. The CaF$_2$ data were hence multiplied with a rising Gaussian function to remove the inhomogeneous broadening. Although this procedure increases the noise level towards the end of the FID, this is a very minor effect with a correction of only {\num{8}} \% for the data point at {\SI{250}{\micro\second}}; hence we did not apply any filter to correct for it. Since the linewidth hampers a precise determination of the exact resonance frequency by means of a simple Fourier transform, the actual offset was determined by analyzing the time-dependence of the phase $\phi(t)$ of the complex valued FID. The offset of \SI{700}{Hz} was removed from the FID. Phase changes due to the variable attenuator were taken into account.

The resulting signal is shown in Fig. \ref{fig:fid}, linear (a) and logarithmic scale (b). The time origin of the FID was assigned to the center of the $\pi/2$ pulse.
The large dynamic range of the FID is only seen in the logarithmic plot in Fig. \ref{fig:fid}(b), where we show the magnitude of the NMR signal's real part. 

\begin{figure*}[t]
  \includegraphics{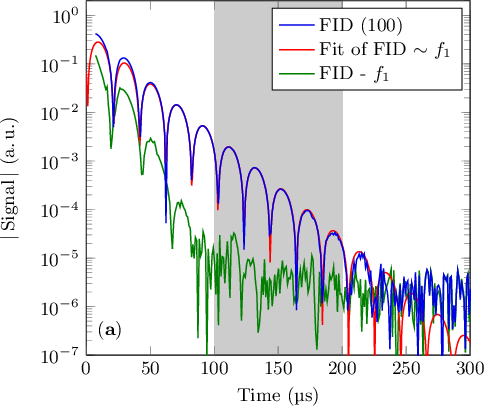}\hfill\includegraphics{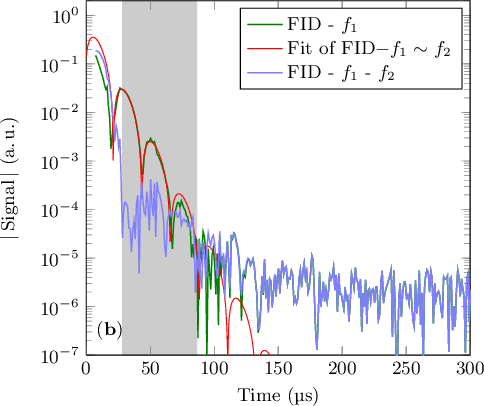}
  \caption{The same plot as in Fig.~\ref{fig:fidTwoModes} but for $B$ parallel to $\left ( 100 \right )$. The FID (\textbf{a}) is again measured with different attenuations; the first \SI{23}{\micro\second} are measured using a spin locking sequence, the time intervals 24 to \SI{51}{\micro\second} and 52 to \SI{111}{\micro\second} are measured with 25 and \SI{12}{dB}, respectively. A Gaussian line narrowing of \SI{426}{\micro\second} was applied to remove the influence of the field inhomogeneity. The fit of the first mode $f_1$ is performed in the time interval 100 to \SI{200}{\micro\second}. The residue, shown as green line in (\textbf{a}), reveals that in this time interval the data is well described by a single mode and again shows oscillatory decaying behavior for earlier times. This oscillatory behavior can well be described in terms of a second mode $f_2$, plotted in (\textbf{b}), with a decay rate about two times the first mode's one.}
  \label{fig:fidTwoModes100}
\end{figure*}

We find good agreement with  Engelsberg and Lowe for times accessible within their experiment, i.e., up to \SI{150}{\micro\second}, c.f. grey curve in Fig.~\ref{fig:fid}(b). The red line in Fig.~\ref{fig:fid}(b) is a fit according to Eq.~(\ref{eq:longTimeTwo}) based on our decay between \num{160} and \SI{250}{\micro\second} using the nonlinear Levenberg-Marquardt algorithm as implemented in Matlab \cite{levenberg_method_1944,marquardt_algorithm_1963}. The algorithm returns estimates, the residuals, the Jacobian and a covariance matrix for the fitted parameters. The 95\% confidence intervals as specified in Tab.~{\ref{tab:fitResults}} are obtained as $a\pm ts$ where $t$ is derived from Student’s $t$ distribution using the degrees of freedom and the required confidence and $s$ is the standard error, i.e., the square root of the corresponding diagonal element of the covariance matrix.  The fit window was chosen since we may expect a  faster second mode and since the algorithm should not operate in low signal-to-noise ratio. We verified that a slight change in the starting point did not change the obtained fit parameters significantly. The obtained values are given in Tab.~\ref{tab:fitResults} as FID (110) $\lambda_1$ (see Tab.~S2 in the supplement {\cite{meier_see_2012}} for all parameters and error estimates). We find a significant deviation between our value ($\gamma_1 = \SI{40.3}{\per\milli\second}$) and the ones determined from Engelsberg and Lowe's data, but also those by Sorte et al. ($\sim \SI{43}{\per\milli\second}$). We also note a substantial difference between the fit (red line) and the data for shorter times. This is shown more clearly in Fig.~\ref{fig:fidTwoModes}(a) where the green curve is the difference between the actual decay and the fit to the long-time decay (first mode $\lambda_1 = \gamma_1 + i \omega_1$). Clearly, the green curve in Fig.~\ref{fig:fidTwoModes}(a) is strong evidence for Fine's second mode, since it is approximately exponential with a decay rate of roughly $2 \gamma_1$ (It disappears in the noise at about \SI{140}{\micro \second} while the first mode has disappeared at \SI{300}{\micro\second}). Note that this result could not be obtained by earlier measurements due to the lower signal-to-noise ratio that did not permit a fit at sufficiently long times. The significantly higher values for $\gamma$ obtained from earlier measurements are due to the influence of the second mode. 

The parameters of the second mode (green curve) in Fig.~\ref{fig:fidTwoModes}(a) are estimated from a fit to Eq.~(\ref{eq:longTimeTwo}) between 75 and \SI{135}{\micro\second}, shown as red curve in Fig.~\ref{fig:fidTwoModes}(b). The blue curve in Fig.~\ref{fig:fidTwoModes}(b) shows the residuum, revealing that the FID can be described accurately in terms of two modes for times greater than \SI{70}{\micro\second}. The estimates for the second mode are given as FID (110) $\lambda_2$ in Tab.~\ref{tab:fitResults}.

\begin{table}[b]
  \begin{tabular}{lrr}
    \toprule[0.3pt]
    Measurement & $\gamma$ (\si{\per\milli\second}) & $\omega$ (\si{\radian\per\milli\second})\\
    \midrule[0.3pt]
    Engelsberg & 43.0(1) & 101.9(5) \\
    Sorte & 43.3(1) & 106.2(1)\\
    FID (110) $\lambda_1$  & 40.3(2) & 103.5(2) \\
    FID (110) $\lambda_2$ & 92(3) & 85(1)\\
    FID (100) $\lambda_1$  & 48.4(3) & 154.0(2) \\
    FID (100) $\lambda_2$ & 109(2) & 142(2)\\
    \bottomrule[0.3pt]
  \end{tabular}
  \caption{Obtained values for $\gamma$ and $\omega$. All values were obtained using a non-linear least squares approximation. For the FID at \SI{20}{K} deviations from the fit are due to thermal noise (c.f. Fig~\ref{fig:fidTwoModes}), hence 95\% confidence intervals were estimated from the covariance matrix.}
  \label{tab:fitResults}
\end{table}

While we are convinced that the green curve in Fig.~\ref{fig:fidTwoModes}(a) is evidence for Fine's second mode, we decided on an independent check, i.e., the measurement of the FID of CaF$_2$ for a different orientation of the crystal with respect to the applied magnetic field. The resulting data and analysis are summarized in Fig.~\ref{fig:fidTwoModes100}. Due to the dependendence of the secular part of the dipolar Hamiltonian on the orientation of the applied magnetic field, different shapes of the FID are obtained for different orientations. For the $(100)$ orientation the FID decays faster and oscillates at a higher frequency. The data can again be understood in terms of two modes with the decay rate of the second mode being about two times larger than the rate of the first mode. The fit results for both modes are given in Tab.~\ref{tab:fitResults}.

To conclude, we have studied the long-time behavior of a macroscopic system of dipole coupled spins $1/2$. The data is interpreted in terms of Fine's theory based on the notion of microscopic chaos. While earlier work is in agreement with this theory, so far only a single decay mode was found. Therefore, it is difficult to discriminate other theories using a memory function approach.
By increasing the signal-to-noise ratio by two orders of magnitude, compared to so far available measurements,  we can determine the first decay mode with high accuracy. After subtracting this mode from the FID we are left with a second decay mode that decays about two times faster than the first mode. Our findings thus support Fine's theory that predicted a well-isolated second decay mode and correctly estimated the difference between the decay rates of the first and the second mode.

As similar accounts in support of Fine's prediction will be given, one can hope that theory will eventually be able to predict the parameters of the observed modes from the Hamiltonian of the system.

\begin{acknowledgments}
  B.M., J.K., and J.H. would like to thank S. Berger, M. Bertmer, M.S. Conradi, C.P. Dietrich (XRD), P. Esquinazi, B.V. Fine, D. Rybicki, H. Voigt.
This work was supported by the DFG within the Graduate School BuildMoNa and ESF project no. 080935191.
\end{acknowledgments}



%

\end{document}